\documentclass[aps,prb,english,superscriptaddress,twocolumn,tightenlines,showpacs,longbibliography,floatfix,amsfonts]{revtex4-2}
\usepackage[T1]{fontenc}
\usepackage[utf8]{inputenc}   
\usepackage[normalem]{ulem}
\usepackage{textcomp}
\usepackage{graphicx,float}
\usepackage[dvipsnames,svgnames]{xcolor}
\usepackage{mathrsfs,mathtools,xfrac}
\usepackage{amsmath,amssymb,amsthm,amscd,bm,bbm}
\usepackage{physics}
\usepackage{txfonts} 
\usepackage{afterpackage,multirow,comment}
\usepackage{array,booktabs,minibox}
	\newcolumntype{P}[1]{>{\centering\arraybackslash}p{#1}} 
\usepackage[colorlinks,bookmarks=false,citecolor=blue,linkcolor=blue,urlcolor=blue,filecolor=blue]{hyperref}
\usepackage{wasysym}
\usepackage[normalem]{ulem}
\theoremstyle{remark}
\newtheorem*{claim*}{\protect\claimname}

\usepackage{babel}
\providecommand{\claimname}{Claim}

\newcommand{\mean}[1]{\left\langle #1 \right\rangle}

\newcolumntype{C}{>{$}c<{$}}
\AtBeginDocument{
\heavyrulewidth=.08em
\lightrulewidth=.05em
\cmidrulewidth=.03em
\belowrulesep=.65ex
\belowbottomsep=0pt
\aboverulesep=.4ex
\abovetopsep=0pt
\cmidrulesep=\doublerulesep
\cmidrulekern=.5em
\defaultaddspace=.5em
}

\begin{document}

\title{Error Resilience of Fracton Codes and Near Saturation of Code-Capacity Threshold in 3$D$}

\author{Giovanni Canossa}
\affiliation{Arnold Sommerfeld Center for Theoretical Physics, University of Munich, 80333 M\"{u}nchen, Germany}
\affiliation{Munich Center for Quantum Science and Technology (MCQST), 80799 M\"{u}nchen, Germany}
\author{Lode Pollet}
\affiliation{Arnold Sommerfeld Center for Theoretical Physics, University of Munich, 80333 M\"{u}nchen, Germany}
\affiliation{Munich Center for Quantum Science and Technology (MCQST), 80799 M\"{u}nchen, Germany}
\author{Miguel A. Martin-Delgado}
\affiliation{Departamento de F\'isica Te\'orica, Universidad Complutense, 28040 Madrid, Spain}
\affiliation{CCS-Center for Computational Simulation, Universidad Polit\'ocnica de Madrid, 28660 Boadilla del Monte, Madrid, Spain}
\author{Hao Song}
\email{songhao@itp.ac.cn}
\affiliation{Institute of Theoretical Physics, Chinese Academy of Sciences, Beijing 100190, China}
\author{Ke Liu}
\email{ke.liu@ustc.edu.cn}
\affiliation{Hefei National Research Center for Physical Sciences at the Microscale and School of Physical Sciences, University of Science and Technology of China, Hefei, 230026, China}
\affiliation{Shanghai Research Center for Quantum Science and CAS Center for Excellence in Quantum Information and Quantum Physics, University of Science and Technology of China, Shanghai, 201315, China}
\affiliation{Arnold Sommerfeld Center for Theoretical Physics, University of Munich, 80333 M\"{u}nchen, Germany}
\affiliation{Munich Center for Quantum Science and Technology (MCQST), 80799 M\"{u}nchen, Germany}
\date{\today}

\begin{abstract} 
Fracton codes have been intensively studied as novel topological states of matter, yet their fault-tolerant properties remain largely unexplored. Here, we investigate the optimal thresholds of self-dual fracton codes, in particular the checkerboard code, against stochastic Pauli noise. By utilizing a statistical-mechanical mapping combined with large-scale parallel tempering Monte Carlo simulations, we calculate the optimal code-capacity threshold of the checkerboard code to be $p_{th} \simeq 0.107(3)$.
This value is the highest among known three-dimensional codes and nearly saturates the theoretical limit for topological codes. Our results further validate the generalized entropy relation for two mutually dual models, $H(p_{th}) + H(\tilde{p}_{th}) \approx 1$, and extend its applicability beyond standard topological codes. This verification indicates the Haah's code also possesses a code capacity threshold near the theoretical limit $p_{th} \approx 0.11$. These findings highlight fracton codes as highly resilient quantum memory and demonstrate the utility of duality techniques in analyzing intricate error-correcting codes.
\end{abstract}

\maketitle

\section{Introduction} \label{sec:intro} 

Quantum error correction (QEC) plays a pivotal role in the development of large-scale quantum computing~\cite{Shor96,Knill97,Gottesman98,Dennis02}. Currently, the leading QEC strategy is the surface code, a two-dimensional topological code that has been realized across various quantum computing platforms~\cite{Google25,He25,Moses23,Bluvstein24}. 
Nevertheless, despite its structural simplicity and ease of experimental implementation, the surface code suffers from significant drawbacks, specifically regarding its inefficiency in encoding logical qubits and performing logical operations~\cite{Pantaleev22,Bravyi13,Pastawski15,Campbell17}.
Consequently, the search for more efficient QEC schemes that surpass the performance of the surface code has become a major focus in the modern study of fault-tolerant quantum computation. Simultaneously, as the construction and fault-tolerance properties of error-correcting codes are deeply rooted in the physics of topological order and statistical mechanics~\cite{Dennis02,Bombin06,Bombin07a,Terhal15,Brown16}, the exploration of new QEC strategies also holds promise for discovering novel phases and new physical phenomena.

Over the past years, a wide array of quantum codes beyond the surface code has been proposed, including higher-dimensional variants of topological codes~\cite{Campbell17} and numerous quantum low-density parity-check codes utilizing remote interactions~\cite{Breuckmann21b}. Nevertheless, in stark contrast to the diversity of the codes, our understanding of their fault-tolerance properties remains rather limited. In particular, the construction primarily focuses on the code parameters $[[n,k,d]]$, where $n$ and $k$ represent the number of physical qubits and logical qubits, respectively, and $d$ denotes the code distance. While these parameters serve as key metrics for a code's capability in encoding logical information, crucial metrics of the code's resilience against errors are still lacking.
Specifically, a code possessing favorable parameters may inherently suffer from a low error threshold, thereby rendering it practically irrelevant. Therefore, determining the fault-tolerance threshold is a critical subsequent step in characterizing these codes. Ideally, one would employ a decoding algorithm capable of handling realistic error models and identify the accurate threshold. However, due to the hypergraph structure inherent in the underlying decoding problem, efficient decoding algorithms remain unknown for the most of those codes. As demonstrated in various notable studies of topological codes~\cite{Dennis02,Ohno2004,Bombin12,Kubica18,Chubb18,Song22}, identifying the accurate threshold can prove highly non-trivial, even under the simplest error models.

Fracton codes constitute a novel class of three-dimensional topological quantum codes that are of significant interest to both condensed matter physics and quantum computing~\cite{Haah11,Vijay16, Vijay15, Bravyi13_RG,Chamon05,Yoshida13,Ma17,Nandkishore19,Song19,WenDN,Aasen20,Pretko20,Zhou22,Song24}. 
In contrast to standard topological codes, which typically possess a constant ground-state degeneracy (GSD), fracton codes are characterized by a sub-extensive GSD and excitations that exhibit restricted mobility or immobility. Notably, previous studies on the X-cube fracton code have demonstrated strong error resilience, evidenced by a remarkably high threshold~\cite{Song22}. Specifically, the X-cube code achieves a code capacity threshold of approximately $0.075$,
considerably surpassing the $0.033$ threshold of the $3D$ toric code~\cite{Ohno2004} and the $0.019$ threshold of the $3D$ color code~\cite{Kubica18}. 
Nonetheless, although fracton codes have been extensively examined through the lenses of condensed matter theory, their fault-tolerant properties remain largely unexplored. To the best of our knowledge, the X-cube code represents the solitary case whose optimal threshold has been explicitly computed.

\begin{table*}[!t]\label{tab:threshold}
\centering
\setlength{\tabcolsep}{24pt}
\renewcommand{\arraystretch}{1.3}
\begin{tabular}{@{}lccc@{}}
\toprule
\hline
Code & $p_{th}^X$ & $p_{th}^Z$ & $H(p_{th}^X) + H(p_{th}^Z)$\\
\hline
$2D$ Surface code
  & $0.1094(2)$~\cite{Dennis02} &  $0.1094(2)$~\cite{Dennis02} & $0.9962(9)$  \\
$2D$ Color code
  & $0.109(2)$~\cite{Katzgraber09} & $0.109(2)$~\cite{Katzgraber09} & $0.994(9)$ \\
$3D$ Toric code
  & $0.2327(3)$~\cite{Ozeki1998} & $0.033(4)$~\cite{Ohno2004} & $0.99(2)$\\
$3D$ Color code
  & $0.276$~\cite{Kubica18} & $0.019$~\cite{Kubica18} & $0.986$ \\
X-Cube code
  & $0.152(4)$~\cite{Song22} & $0.075(2)$~\cite{Song22} & $1.00(1)$\\
{\bf Checkerboard code}
  & ${\bf 0.107(3)}$   & ${\bf 0.107(3)}$ & ${\bf 0.98(2)}$ \\
Haah's code
  & $[\approx 0.11]$   & $[\approx 0.11]$ & $[\approx 1.00]$ \\
\hline
\bottomrule
\end{tabular}
\caption{Optimal code capacity thresholds of representative topological codes.
Results of this work are bold-faced. $[\cdot]$ indicates the expected value from the generalized entropy duality.}
\end{table*}

In this work, we investigate self-dual fracton codes and compute their optimal threshold against stochastic Pauli errors. The determination of the error threshold can be mapped to a phase transition problem, allowing us to determine the optimal threshold value. However, the exact calculation of this optimal threshold is highly non-trivial, as it corresponds to analyzing phase transitions in three-dimensional random spin models which are notoriously difficult to simulate. The complexity is further exacerbated in this context because the relevant spin models involve multi-spin interactions and very strong first-order phase transitions.
This leads to immense demands of computational resources as the calculation for a single code potentially require {\it millions} of CPU hours. Therefore, we restrict our numerical simulations on the checkerboard code, while also analyzing the threshold of Haah's code through a generalized entropy duality.

The importance of our work is twofold. First, we determine that the checkerboard code exhibits a remarkably high code capacity threshold of $p_{th} \simeq 0.107(3)$, even under a conservative estimate. This represents the first instance of a three-dimensional code nearly approaching the saturation value of $\approx 0.11$. While our analysis is restricted to random Pauli noise, this simplified error model constitutes a fundamental starting point and is extensively employed in the study of codes with complicated structures. Moreover, our result also establishes a concrete theoretical benchmark to guide the future development of new efficient decoders.

Second, we corroborate the validity of a powerful generalized entropy duality across a broader scope of topological codes. 
It is expected that the $X$ and $Z$ code-capacity thresholds of a Calderbank–Shor–Steane (CSS) code with a zero rate satisfy the inequality $H(p_{th}^X) + H(p_{th}^Z) \leq 1$, where $H(p) := -p \log_2 p - (1-p) \log_2 (1-p)$ denotes the binary Shannon entropy~\cite{Dennis02,Gottesman97}. From the perspective of statistical physics, the saturation of this bound corresponds to the duality relation $H(p_{th}) + H(\tilde{p}_{th}) \approx 1$, where $p_{th}$ and $\tilde{p}_{th}$ denote the critical points of two mutually dual models~\cite{Nishimori07,Song22}. This implies that determining the critical point for one model allows for the deduction of the critical point for its dual. And for self-dual models, the critical point can be immediately estimated.
Moreover, while this work focuses on code capacity thresholds, this duality extends to include scenarios involving measurement errors~\cite{Wang03,Xu25}.

It is important to note that, although this generalized duality can save significant computational resources, it is rooted in a self-consistent replica-symmetry-breaking approximation~\cite{Nishimori07,Song22}. Therefore, it must be sufficiently verified before it can be confidently applied to different codes. Table~\ref{tab:threshold} summarizes the success of this duality to standard topological codes. Combined with our current and previous findings, we confirm its applicability to fracton codes. Supported by these extensive justifications, we expect that Haah's code, whose simulation is even more resource demanding, also nearly saturates the code capacity threshold  $p_{th} \approx 0.11$.

The manuscript is organized as follows. Section~\ref{sec:Checkerboard} describes the checkerboard code and its key properties. Section~\ref{sec:mapping} is dedicated to a brief overview of the statistical-mechanical mapping and the generalized entropy duality relation.
The problem of optimal error threshold is also discussed.
Section~\ref{sec:results} presents an analytical statistical-mechanical analysis of the checkerboard code and our numerical results.
Section~\ref{sec:Haah} discusses the situation of the Haah's code.
Finally, we conclude in Section~\ref{sec:sum}.

\section{Checkerboard code} \label{sec:Checkerboard}
\begin{figure}
\includegraphics[width =.48\textwidth]{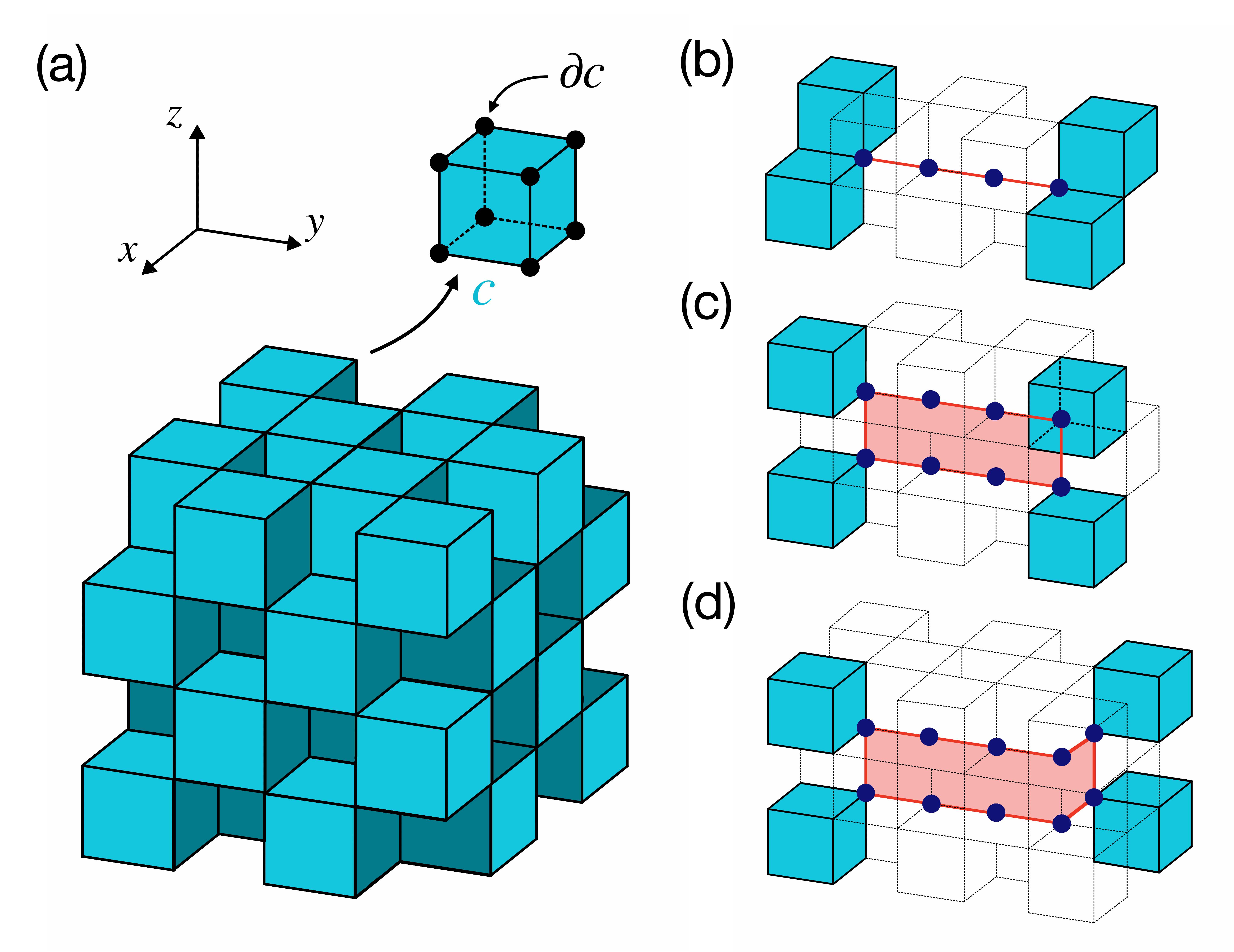}
\caption{Illustration of the checkerboard code and its excitations. (a) Physical qubits (black dots) reside on the vertices of the lattice. Stabilizers $\hat{A}_c$ and $\hat{B}_c$ are defined on each colored cube $c$ as products of $\hat{\sigma}^x$ and $\hat{\sigma}^z$ operators, respectively, acting on the vertex set $\partial c$.
(b) Fracton excitations are generated in quartets by a rigid one-dimensional string operator.
(c, d) A dipolar pair of fractons with even separation moves freely in the plane perpendicular to its dipole moment.
}
\label{fig:model}
\end{figure}

The checkerboard code is defined on a cubic lattice of linear size $L$ with periodic boundary conditions (PBC), where a single qubit resides at each lattice site.
The stabilizer operators $\hat{A}_c=\prod_{i\in \partial c} \hat{\sigma}_i^{x}$ and $\hat{B}_c=\prod_{i\in \partial c} \hat{\sigma}_i^{z}$ are defined at the center of every other cube $c$ as the tensor product of Pauli $\hat{\sigma}^x$ and $\hat{\sigma}^z$ operators between the 8 spins residing at its vertices. 
These stabilizers are distributed throughout the lattice in a checkerboard pattern as shown in Fig.~\ref{fig:model}(a), so that in a system of linear size $L$ there are $L^3/2$ $\hat{A}_c$ stabilizers and $L^3/2$ $\hat{B}_c$ stabilizers. To satisfy both periodic boundary conditions (PBC) and the checkerboard structure, the model is defined only for even $L$. 
These conditions ensure that neighbouring stabilizers always share 2 qubits, guaranteeing that all stabilizers commute with each other and that the energy spectrum and ground state degeneracy can be determined analytically, with the ground states $\ket{\psi}$ satisfying the condition $\hat{A}_c\ket{\psi}=\hat{B}_c\ket{\psi}=+\ket{\psi}\,\,  \forall c$.

Since the model is invariant under the exchange $\hat{\sigma}^x \leftrightarrow \hat{\sigma}^z$, the $\hat{A}_c$ and $\hat{B}_c$ stabilizers share the same excitation spectrum and dynamics. The model is characterized by a $\mathbb{Z}_2$ charge conservation law for each $IJ$-plane of cubes ($IJ=xy,yz,zx$), as indicated by the relation $\prod_{c\in IJ-\text{plane}} \hat{A}_c =\prod_{c\in IJ-\text{plane}} \hat{B}_c =1$ which enforces that any excitation above the ground state must satisfy these $\mathbb{Z}_2$ conservation laws on each plane. Due to the geometry of the checkerboard distribution, the application of an individual $\hat{\sigma}^x$ or $\hat{\sigma}^z$ operator to a specific qubit generates four elementary excitations, known as fractons, associated with the $\hat{B}_c$ or $\hat{A}_c$ stabilizers sharing the affected qubit.

These fractons can be viewed as gapped topological defects generated on top of the ground state. Due to overlapping $\mathbb{Z}_2$ charge conservation laws, \cal{individual fractons cannot be moved without creating additional excitations}. However, a pair of fractons (a dipole) possesses restricted mobility in directions orthogonal to its dipole moment. For example, as depicted in Fig.~\ref{fig:model}(b), an adjacent pair of fractons (a dipole) living in one $xz$-plane can be moved without energy cost along a rigid line $\textit{l}$ in the $y$-direction by applying string operator $\prod_{n\in \textit{l}}\hat{\sigma}^x_n$ or $\prod_{n\in \textit{l}}\hat{\sigma}^{x/z}_n$. Moreover, if a dipole is oriented along the z-axis, the pair can move freely along the $x$ and $y$ directions. This motion is generated by stacks of adjacent $z$-strings, forming a flexible membrane that permits motion within the $xy$-plane, as shown in Fig.~\ref{fig:model}(c, d).

Due to the planar conservation laws and the redundancies introduced by PBC, the checkerboard code on a lattice of $L^3$ qubits exhibits a total ground-state degeneracy of $2^{3L-6}$. Thus its ground-state manifold can be regarded as a computational space comprising $3L-6$ logical qubits, each associated with a pair of anticommuting logical operators $X_L$ and $Z_L$.

\section{Statistical Mapping and Duality} \label{sec:mapping}

\subsection{Error Correction} \label{sec:errorcorrection} 
For a given stabilizer model, let \(Q\) denote the set of physical qubits. The commuting $X$-type and $Z$-type stabilizer operators are denoted by $\hat{A}_{\mathcal{S}_X}$ and $\hat{B}_{\mathcal{S}_Z}$, respectively, where $\mathcal{S}_{X/Z}$ serves as both a label for the stabilizer and an identifier for its anchor cell (in the Checkerboard model, these would be the cubic cells $c$ as shown in Fig.~\ref{fig:model}). The supports of these operators on the lattice are denoted by $\partial \mathcal{S}_X, \partial \mathcal{S}_Z \subset Q$. The logical code space is defined as the ground-state subspace of the Hamiltonian:
\begin{align}\label{eq:Check_Ham}
&\mathcal{H}= - \sum_{\mathcal{S}_X} \hat{A}_{\mathcal{S}_X}
   \;-\; \sum_{\mathcal{S}_Z} \hat{B}_{\mathcal{S}_Z},\\
&\text{where }\,\,\hat{A}_{\mathcal{S}_X} = \prod_{i \in \partial \mathcal{S}_X} \hat{\sigma}_i^{x}, \qquad \hat{B}_{\mathcal{S}_Z} = \prod_{i \in \partial \mathcal{S}_Z} \hat{\sigma}_i^{z}. \nonumber
\end{align}

Error correction in topological stabilizer codes proceeds by identifying the error configuration $\eta$, which can be represented as a product of Pauli $X$ and $Z$ operators acting on the qubits. When the system is initialized in a ground state of $\mathcal{H}$, an error generally drives it into an excited state in which a subset of the stabilizers acquires eigenvalue $-1$. This pattern of violated stabilizers defines the \emph{error syndrome} $s$.

Because the exact configuration $\eta$ is typically unknown, the syndrome $s$ is used to estimate the most likely logical error class. Two error configurations $\eta$ and $\eta'$ are considered equivalent if they differ only by a product of stabilizers $\hat{\mathcal{O}}$ (i.e., $\eta' = \hat{\mathcal{O}}\eta$). This equivalence breaks down if the configurations differ by a logical operator $\lambda$. Consequently, for a given syndrome $s$, errors are grouped into equivalence classes $[\eta_s + \lambda]$. The probability of a syndrome is given by the sum over these classes: $\mathrm{Pr}(s) = \sum_{\lambda} \mathrm{Pr}([\eta_s + \lambda])$, and the most likely logical error class is the one with the highest total probability.

Error correction is asymptotically reliable if there exists a unique most-probable class $[\eta_s^*]$ such that, in the thermodynamic limit, $\mathrm{Pr}([\eta_s^*]) \to \mathrm{Pr}(s)$ \cite{Dennis02}. This condition holds only when physical error rates remain below a threshold probability $p_{\text{th}}$, a limit guaranteed by the threshold theorem \cite{Aharanov99}. Above this threshold, the logical error class cannot be reliably determined; thus, the optimal threshold is a fundamental parameter for assessing the performance of any quantum code.

In this work, we focus on independent noise models where $X$ and $Z$ error syndromes are decoupled and treated separately. For self-dual models, such as the checkerboard code and Haah's cubic code, the error analysis for $X$ and $Z$ sectors is identical, leading to equal threshold probabilities: $p_{\text{th}}^X = p_{\text{th}}^Z \equiv p_{\text{th}}$.

\subsection{Statistical Mechanical Mapping}\label{sec:statmech}
To ascertain whether it is possible to identify the most likely error class for a given physical error probability $p$, one can relate the quantum information properties of a given error model to the physical features of a classical Ising spin model subject to disorder. Ever since its first introduction in Ref.~\cite{Dennis02}, this relation has been an active subject of study and has been formalized under the name of the statistical-mechanical (SM) mapping~\cite{Chubb18}.

Focusing on a noise model with independent $X$ (or $Z$) errors - whose statistics can be treated separately - consider an error configuration $\eta\subseteq Q$ (i.e. a set of qubits affected by an error) that is consistent with a given observed syndrome $s$. For any other error configuration $\eta'=\eta + V$ obtained by further applying errors to the set of qubits $V\subseteq Q$, we have
\begin{flalign}\label{stat1}
P(\eta + V)& =P(\eta)\,\frac{P(\eta+V)}{P(\eta)}  \nonumber \\
&= P(\eta)\prod_{\ell\in V\cap\eta}\frac{1-p}{p}\;
\prod_{\ell\in V\setminus\eta}\frac{p}{1-p},
\end{flalign}
where $\ell$ labels individual qubits, and the products run over those qubits in the specified subsets of $Q$. 
Two configurations belong to the same equivalence class iff they differ by a product of stabilizers $\hat{A}_{\mathcal{S}_X}$ (or $\hat{B}_{\mathcal{S}_Z}$): in that case, the possible flip set $V$  corresponds to the symmetric difference of the supports $\partial \mathcal{S}_X$ (or $\partial \mathcal{S}_Z$). This allows us to express the probability distribution of all error configurations belonging to the same equivalence class through a statistical mechanical partition function of a classical spin model. 

To see this, we start off by placing an Ising degree of freedom $\sigma_{\mathcal{S}_{X/Z}}=\pm 1$ at the center of each stabilizer $\mathcal{S}_{X/Z}$. Then, for each physical qubit error $X_i(Z_i)$, define the set of stabilizers that share it as $\mathcal N_{X(Z)}(i)\equiv \{\mathcal{S}_{X(Z)} \,|\,i\in \partial{\mathcal{S}_{X(Z)}}\}$ and introduce a coupling between all Ising spins $\sigma_{\mathcal{S}_{X/Z}}\in\mathcal{N}_{X(Z)}$ corresponding to stabilizers that share said physical qubit operator. For a given initial error configuration $\eta$, the total probabilities of all configurations in the same equivalence class are captured by the following partition function:
\begin{equation}\label{StatMechPartition}
\mathcal{Z}(\eta,T)\,=\, \sum_{\{\sigma\}} \text{exp}\left[ \beta \sum_{i \in Q} \eta_i \prod_{j\in \mathcal{N}_{X(Z)}}\sigma_j \right].
\end{equation}
The coupling strengths $\eta_i=\pm 1$ depend on whether the qubit $i$ was part of the original error configuration ($\eta_i=-1$) or not ($\eta_i=+1$).
The proposed classical model accurately captures the statistical properties of the error model given by Eq.~\eqref{stat1} once we relate the effective inverse temperature $\beta$ to the single-qubit error probability $p$ via the Nishimori condition:
\begin{equation} \label{eq:NL}
e^{-2\beta_N} = \frac{p}{1-p}. 
\end{equation} 
It can be shown that the partition function $\mathcal{Z}(\eta,T)$ depends only on the error equivalence class of $\eta$, consistent with its correspondence to $\mathrm{Pr}([\eta])$.

\subsection{Optimal Error Threshold} \label{sec:OptimalErrorThreshold} 
Given an error syndrome $s$ and a candidate error configuration $\eta$, one can infer the most likely logical error by comparing the relative probabilities of different error classes. For an independent $X$ (or $Z$) error model with a homogeneous error rate $p$, the probability ratio of two equivalence classes distinguished by a logical operator $\lambda$ is reflected by the ratio of the partition functions of the model with error configuration $\eta$ and $\eta+\lambda$, which is equivalent to the free energy cost of applying $\lambda$ onto the system:
\begin{equation}
\frac{\mathrm{Pr}\left[\eta+\lambda\right]_{\left( X(Z),p \right)}}{\mathrm{Pr}\left[\eta\right]_{\left( X(Z),p\right)}} = \frac{\mathcal{Z}(\eta,\beta_N)}{\mathcal{Z}(\eta+\lambda,\beta_N)} \equiv e^{-\beta_N \delta F_{\eta,\lambda}(\beta_N)}.
\end{equation}
To determine the appropriate correction, it is necessary to identify the most probable equivalence class $[\eta_{s}^{*}]_{X(Z)}$ such that, in the thermodynamic limit, $\Pr\left([\eta_{s}^{*}]_{X(Z)}\right) \to \Pr(s)$. For this to occur, the free energy difference $\delta F_{\eta,\lambda}$ associated with the addition of a logical operator must diverge in the thermodynamic limit. Since applying a logical operator corresponds to flipping a set of couplings across the entire lattice of the associated Ising model, the condition $\delta F_{\eta,\lambda} \to \infty$ implies that introducing a domain wall incurs a free energy cost proportional to its size. This condition is met only if the Hamiltonian $H_{\eta}$ supports an ordered phase at finite temperature~\cite{Dennis02}. The existence of an ordered phase at the temperature $\beta_N$ corresponding to the error probability $p$ ensures that, in the thermodynamic limit, the error model can always be decoded for error rates below the threshold. Thus, the optimal error threshold can be inferred by studying the order-disorder phase transition of the classical Ising model with quenched disorder.

\subsection{Generalized Duality in classical spin systems}
As explained in the previous section, the SM mapping allows one to capture the statistics of the possible equivalent error configurations that match a given error syndrome with the partition function of a classical Ising model. 

While the disordered Ising model is characterized by the interplay of disorder and temperature, it accurately describes the statistics of the quantum spin system only when it satisfies the Nishimori condition $e^{-2\beta_N}= p/(1-p)$: this identifies the Nishimori line (NL), a special locus in parameter space along which the quenched disorder distribution is exactly matched to the Boltzmann weight of the classical model~\cite{Nishimori07}. Along the NL, the probability of flipping an Ising spin, dictated by its Boltzmann weight $e^{-\beta E}$, is identical to the probability of flipping the bonds which share said spin (which corresponds to  changing the error configuration, dictated by the error probability $p$). 


The \textit{enhanced symmetry} between thermal and disorder fluctuations renders the Nishimori line a particularly interesting region of parameter space: beyond its already mentioned relation with the statistics of the corresponding QEC error model, this symmetry strongly suggests that the multicritical point of the disordered Ising model lies on the NL, owing to its invariance under renormalization-group transformations~\cite{Doussal89,Iba99}. This property implies the existence of a transition from an error correctable to a non-error correctable phase in the thermodynamic limit \cite{Dennis02,Chubb18}, which is guaranteed by the threshold theorem \cite{Aharanov99}. On top of that, it is exactly along this line that it is possible to define an approximate fixed point condition which relates the position of the multicritical points of any Ising model with its geometrically dual model \cite{Takeda05}. This can be derived by recalling the relation between the partition functions of the model and that of its dual.

To do this, consider a $d$-dimensional Ising Hamiltonian that can be written in the following form:
\begin{equation}
H = -J \sum_{C_j} \prod_{i\in\partial C_j} \sigma_i.
\end{equation}
Here, $C_j$ labels the $r$-dimensional coupling object (i.e., a simplex) which represents an interaction involving all Ising spins $\sigma_i$ lying on its boundary $\partial C_j$ (e.g., a segment on the $2D$ nearest-neighbour Ising model or a tetrahedron in the $3D$ Tetrahedral Ising model). Each of these elements $C$ will contribute to the partition function of $K \equiv \beta J$ with the aligned and anti-aligned Boltzmann factors $u_{\pm}(K) = e^{\pm K}$ which appear in the configurations with $\prod_{i\in\partial C_j} \sigma_i = \pm 1$. 

The geometrically dual model is constructed using the procedure outlined in \cite{Wegner71} and consists of a set of dual Ising spins $\sigma^*$ connected by a set of dual couplings $C^*$. We can introduce the dual weights associated to each coupling of the dual model as the 2-component Fourier transform of the original weights $u^*_{\pm}(K) = \left( u_+ (K) \pm u_{-}(K) \right) /\sqrt{2}$. Under this construction, the partition function of the original and dual lattice are related by the Normal Factor Graph (NFG) Duality theorem \cite{Forney11} which, in this context, represents a generalized version of the Kramers-Wannier Duality \cite{Kramers41a,Savit80}:
\begin{equation}\label{eq:genKW}
\mathbf{Z}\{u_{\pm}(K)\} = 2^a \,\tilde{\mathbf{Z}}\{u^*_{\pm}(K)\}
\end{equation}
where $a$ is some appropriate factor. It follows that the product $\mathbf{Z}\{u_{\pm}(K_1)\}\tilde{\mathbf{Z}}\{u^*_{\pm}(K_2)\}$ is invariant under the simultaneous transformations $u_{\pm}(K_1)\rightleftharpoons u_{\pm}^*(K_2)\,,\,\,u_{\pm}(K_2)\rightleftharpoons u_{\pm}^*(K_1)$: the transition points $K_1,K_2$ of the two models are then given by the fixed point condition~\cite{Nishimori10}
\begin{equation}
u_{\pm}(K^C_1)u_{\pm}(K^C_2)=u_{\pm}^*(K^C_1)u^*_{\pm}(K^C_2),
\end{equation}
which reflects the symmetry under duality transformation.

One can make a similar conclusion in presence of quenched disorder using the replica trick. Consider $n$ replicas of the disordered model characterized by the same Hamiltonian
\begin{equation}
H = -J \sum_{C_j} \eta_j \prod_{i\in\partial C_j} \sigma_i,
\end{equation}
where, $\eta_j=-1$ with probability $p$. One can define an averaged Boltzmann weight for each of the cases where the spins of individual coupling $\partial C$ are antialigned for $0\leq k \leq n$ of the $n$ replicas and are aligned in the remaining $n-k$ replicas. Averaging over the possible values of $\eta_C$, one gets:
\begin{flalign}
x_k(p,K) & =\langle \,\,\text{exp}\left[ K \eta_C  \sum_{\alpha=1}^n \prod_{i\in\partial C}\sigma_i^{\alpha}\right]\,\,\rangle \nonumber\\
& = p e^{(n-2k)K} +(1-p) e^{-(n-2k)K} \nonumber \\
& = p u_+^{n-k}u_-^{k} + (1-p) w_+^{n-k}w_-^{k}
\end{flalign}
where we introduced the index $\alpha$ to distinguish  different replicas, reused the short-hand notation $u_{\pm}(K)$ for the Boltzmann factors in the replicas with $\eta_C=+1$ and introduced $w_{\pm}(K)\equiv e^{\mp K}$ for the replicas with $\eta_C=-1$. 
To take into account all possible spin configuration of the $n$-replicated systems, the disorder-averaged partition function will be a function of all these Boltzmann factors:
\begin{equation}
\langle\mathbf{Z}^n\rangle_{\text{dis}} \equiv \mathbf{Z}_n\{x_0 (p,K),x_1 (p,K), \,...\,x_n (p,K) \}.
\end{equation}
Keeping in mind that $u_{\pm}$ and $w_{\pm}$ correspond to different bond configurations and therefore undergo different Fourier transformations, we can define the dual Boltzmann factors on the dual lattice as the $n$-fold 2-component Fourier transform,
\begin{flalign}\label{eq:dualBoltzmann}
x^*_{k}(p,K) &= \mathcal{F}\left( p u_+^{n-k}u_-^{k} + (1-p) w_+^{n-k}w_-^{k} \right) \\
&= p (u_+^*)^{n-k} (u_-^*)^{k} + (1-p) (w_+^*)^{n-k}(w_-^*)^{k}   \nonumber \\
&= 2^{-n/2}\left(p + (-1)^k(1-p)\right)\left(e^K+e^{-K}\right)^{n-k}\left(e^K-e^{-K}\right)^{k}. \nonumber 
\end{flalign}
the disorder generalization of Eq. \eqref{eq:genKW} directly follows:
\begin{equation}
\mathbf{Z}_n\{x_0,x_1, \,...\,x_n \} = 2^{\bar{a}}\mathbf{Z}_n\{x_0^*,x_1^*, \,...\,x_n^* \}.
\end{equation}
Since the multicritical point is expected to lie along the Nishimori line (Eq.~\eqref{eq:NL}), where the enhanced symmetry reduces the fixed-point conditions to a one-parameter problem, 
the replicated local weights $\{x_k\}$ cannot, in general, satisfy the full set of self-duality equations $x_k=x_k^{*}$ simultaneously in the presence of disorder, as the resulting system of equations is overconstrained.

Restricting our analysis to the Nishimori line, which in our generalized formulation is given by the condition $e^{-2K_N}= p/(1-p)$, Eq.~\eqref{eq:dualBoltzmann} yields an exact hierarchy among the dual Boltzmann factors, distinguished by
the parity of $k$:
\begin{equation}
\frac{x_{k}^*}{x_0^*}=(1-2p)^{k}\,\,\text{(even)},\qquad
\frac{x_{k}^*}{x_0^*}=-(1-2p)^{k+1}\,\,\text{(odd)}.
\end{equation}
For $0<p<1$, the $k>0$ sectors are suppressed in magnitude relative to the principal factor, with $|x_{k}^*/x_0^*|\leq(1-2p)^2<1$. Following the standard replica approach, we set $K=K_N(p)$ and impose the \emph{principal-factor} condition only on the $k=0$ sector. This gives ~\cite{Takeda05,Nishimori07,Song22}
\begin{equation}
x_0(p_1^C,K_1^C)x_0(p_2^C,K_2^C) \approx x_0^*(p_1^C,K_1^C)x_0^*(p_2^C,K_2^C).
\end{equation}
The inclusion of subleading sectors ($k>0$) would shift the estimate slightly while leaving the leading condition as the dominant constraint. Nevertheless, it was shown in several cases to yield the closest approximation of the true multicritical point \cite{Hinczewski05,Hasenbusch07,Masayuki09,Kubica18,Gattringer18,Song22}. Employing the replica trick limit $n\rightarrow0$, the principal factor condition reduces to
\begin{equation}\label{eq:duality}
H(p_1^C) + H(p_2^C) \approx 1,
\end{equation}
where $H(p) := -p \,\text{log} p - (1-p)\,\text{log} (1-p)$.
This corresponds to the limit of the bound
$H(p_{th}^X) + H(p_{th}^Z) \leq 1$.
While the derivation of Eq.~\eqref{eq:duality} relies on the leading Boltzmann factor, i.e., on the replica analysis, it works surprisingly well for known topological codes as summarized in Table~\ref{tab:threshold}.

With this in mind, we turn our attention to the $3D$ fracton models to determine whether the presence of subsystem symmetries may affect the reliability of this prediction. 
Specifically, as the non-self-dual X-cube code was shown to saturate the error threshold bound in a previous work \cite{Song22}, here we complement it for the self-dual checkerboard code.

\section{Effective spin model and Numerical simulations} \label{sec:results}

\subsection{Random Spin Models With Subsystem Symmetry}
In the checkerboard code, each qubit is shared by four stabilizers; using the method outlined in section~\ref{sec:statmech}, the error model can be mapped into a random Tetrahedral Ising model~\cite{Canossa24,Vijay15}. For both the $X$ and $Z$ error model, each qubit $i$ lies at the corner of 4 stabilizers of the checkerboard code: in the effective spin model, we will label the set of Ising degrees of freedom associated to the stabilizer sharing said qubit $i$ as \{$\sigma_j$,$\sigma_k$,$\sigma_l$,$\sigma_m$\}=$\mathcal{N}_{X}(i)=\mathcal{N}_{Z}(i)$. 

\begin{figure}[!t]
\centering
\includegraphics[width =.48\textwidth]{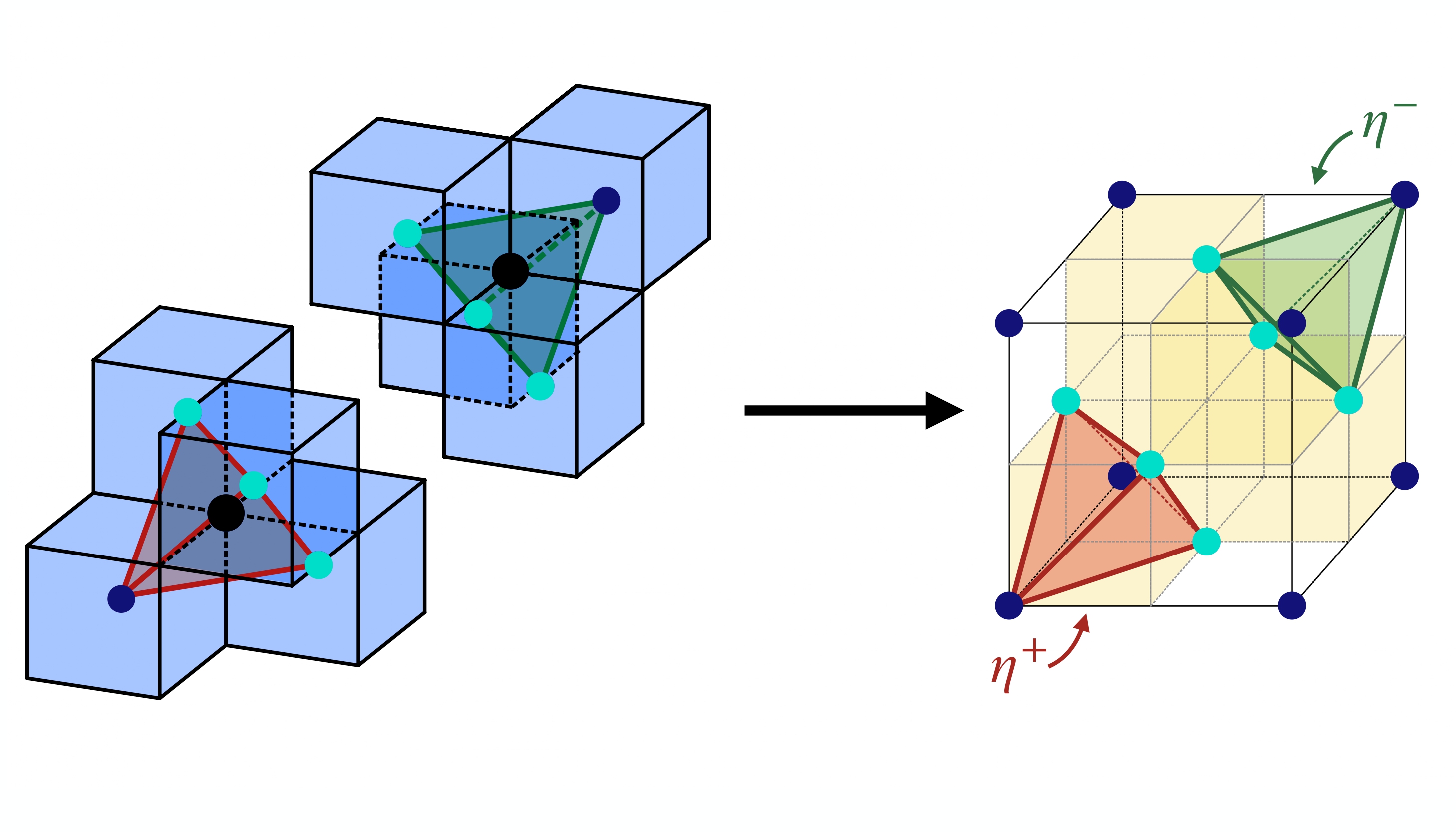}
\caption{Illustration of the statistical mapping from the checkerboard code (left) to the Tetrahedral Ising model on a face-centered cubic lattice (right) derived from~\cite{Canossa24}. Blue and cyan circles, located at the vertices and face centers of the FCC lattice, represent the Ising spins associated with each stabilizer, constructing the dual lattice of the checkerboard code. The action of a single qubit error is mapped into a four-body interaction with strength $\eta_{\mathbf{v}}^{\pm}$; these are defined for each lattice site $\mathbf{v}$ in the FCC lattice and are represented by a red (green) tetrahedron placed in each shaded cube (blank cube) of the lattice on the right. }
\label{fig:Tetra_mapping}
\end{figure}

The partition function in Eq.~\eqref{StatMechPartition} can be rewritten as
\begin{equation}
\mathcal{Z}(\eta,T) = \sum_{\{\sigma\}}
\text{exp}\left[\beta \sum_{i\in Q } \eta_i\,\, \sigma_j \sigma_k \sigma_l \sigma_{m} \right].
\end{equation}
This can be rewritten in a more convenient form on a face-centered cubic lattice comprising $\frac{1}{2}L^3$ spins for a system of linear size $L$ as shown in Fig.~\ref{fig:Tetra_mapping}. Labelling the lattice sites as $\mathbf{v} = (x,y,z)\in \mathbb{Z}^3$, the Hamiltonian of the model is given by
\begin{flalign}
\mathcal{H} = -\sum_{\mathbf{v}}\Big( \,\,& \eta^{+}_{\mathbf{v}}\,\sigma_{\mathbf{v}}\sigma_{\mathbf{v}+\hat{x}+\hat{y}}\sigma_{\mathbf{v}+\hat{y}+\hat{z}}\sigma_{\mathbf{v}+\hat{x}+\hat{z}} \nonumber \\ +  & \eta^{-}_{\mathbf{v}}\,\sigma_{\mathbf{v}}\sigma_{\mathbf{v}-\hat{x}-\hat{y}}\sigma_{\mathbf{v}-\hat{y}-\hat{z}}\sigma_{\mathbf{v}-\hat{x}-\hat{z}} \,\,\Big) 
\end{flalign}
where $\eta^{\pm}_{\mathbf{v}}$ is labelling the upwards and downwards 4-body tetrahedra $\eta_{i}$.\\

Beyond the trivial global $\mathbb{Z}_2$ symmetry, the Hamiltonian is invariant under flipping all spins in any arbitrary $XY$, $YZ$, or $XZ$ layer of the lattice. This subextensive set of subsystem symmetries undergoes spontaneous symmetry breaking, leading to a subextensive ground-state degeneracy of $2^{3L-6}$. This degeneracy directly reflects the ability of the checkerboard code to encode $3L-6$ logical qubits.

\subsection{Disorder Simulations} \label{app:simulations}
\begin{figure}[t]
\centering
\includegraphics[width =.48\textwidth]{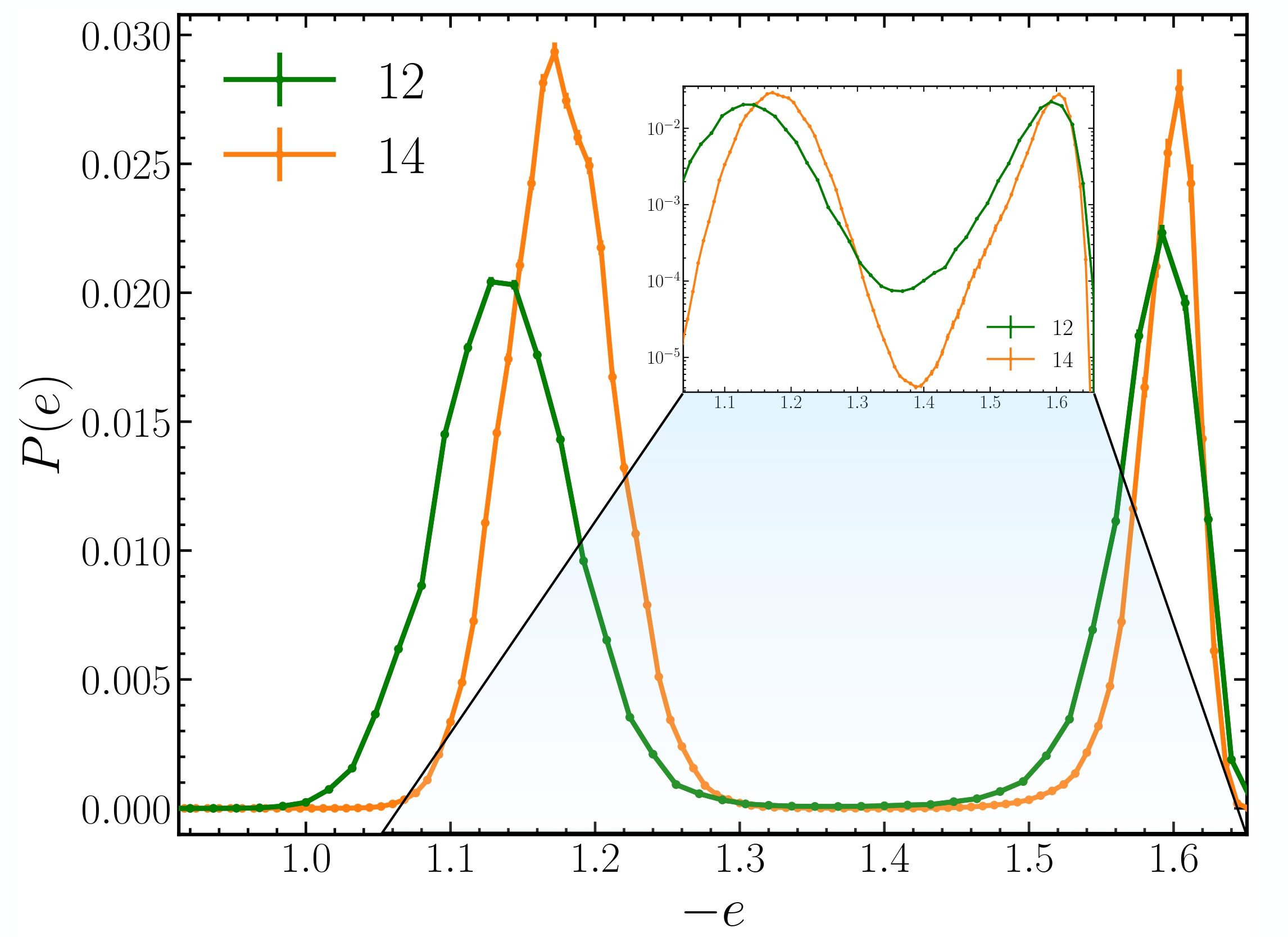}
\caption{Energy histograms at the transition temperature for a representative disorder value below threshold ($p = 0.090$). While a sharpening of the double-peak structure with increasing system size is consistent with a first-order transition, it can also occur if the two peaks move closer together as $L$ grows, reflecting strong finite-size effects. The decisive signature of a first-order transition is the suppression of energetic configurations between the peaks, which is expected to become more pronounced with increasing system size (as shown in the inset). This provides a reference for the expected behavior below the error threshold.}
\label{fig:090}
\end{figure}

The Tetrahedral Ising model with no disorder is characterized by a very strong first-order phase transition~\cite{Canossa24}. While quenched disorder reduces the amount of latent heat required to break the phase boundaries, the features of the first-order transition are still relatively sharp near the threshold (see Fig.~\ref{fig:090}): this leads to very slow equilibration which makes the simulation of each disorder configuration of the error model severely expensive from a computational perspective. When paired to the fact that one needs to study the average of data from several hundreds of disorder iterations in order to study the disorder average, it is necessary to optimize the individual simulations in order to minimize their equilibration time and overall runtime. With this in mind, each disorder realization has been simulated with parallel tempering in conjunction with Metropolis-Hastings and overrelaxation updates in order to speed up the equilibration process~\cite{Katzgraber06, Janke08}. The temperature distribution is tailored for all disorder simulations at a given disorder value $p$ and system size $L$ in order to maximize the mixing of replicas between ordered and disordered phase in the average case.

The disorder averages are obtained through a large number of configurations ($N_d \sim 500$), and their statistical uncertainties are derived through the Jackknife resampling method~\cite{Janke08}. The specific number of temperatures and disorder iterations used for each disorder average is reported in Table~\ref{tab:simparams}. 

Equilibration of each individual disorder iteration is inferred by combining various analyses. The timeseries of the energy is divided into bins with intervals $[2^{\tau}, 2^{\tau+1}-1]$, where $\tau\in\mathbb{N}$ labels a specific bin. After averaging the values stored in each bin, the system can be considered equilibrated when the last values of the last three bins are within error bars of each other~\cite{Bombin12}.

While this method ensures equilibration at low temperatures, it is less reliable when studying temperatures near the phase transition, where the system visits both phases over time. In most cases, a qualitative analysis of the $\chi_O$, $C_V$, and $O$ was sufficient to infer whether a disorder simulation had failed to reach equilibration or if too few samples were taken. 

In the few cases where this was not sufficient, the energy histograms were tested through a reweighting method~\cite{Janke08}: after collecting the energy histogram for each inverse temperature $\beta$, the energy histogram at a new temperature $\beta'$ can be derived by applying to each bin at total energy $E$ a weight $e^{-(\beta'-\beta)E}$. The simulation is considered equilibrated if, for most temperatures near the phase transition, the energy histograms agree within sufficient statistical uncertainty after appropriate reweighting. This ensures that, while the disorder average at certain individual temperatures might still have gotten stuck in one of the two phases, the majority of the replicas have consistently captured the expected behaviour of the system at their respective temperatures.

\begin{figure*}[!tb]
\centering
  \includegraphics[width=1.0\textwidth]{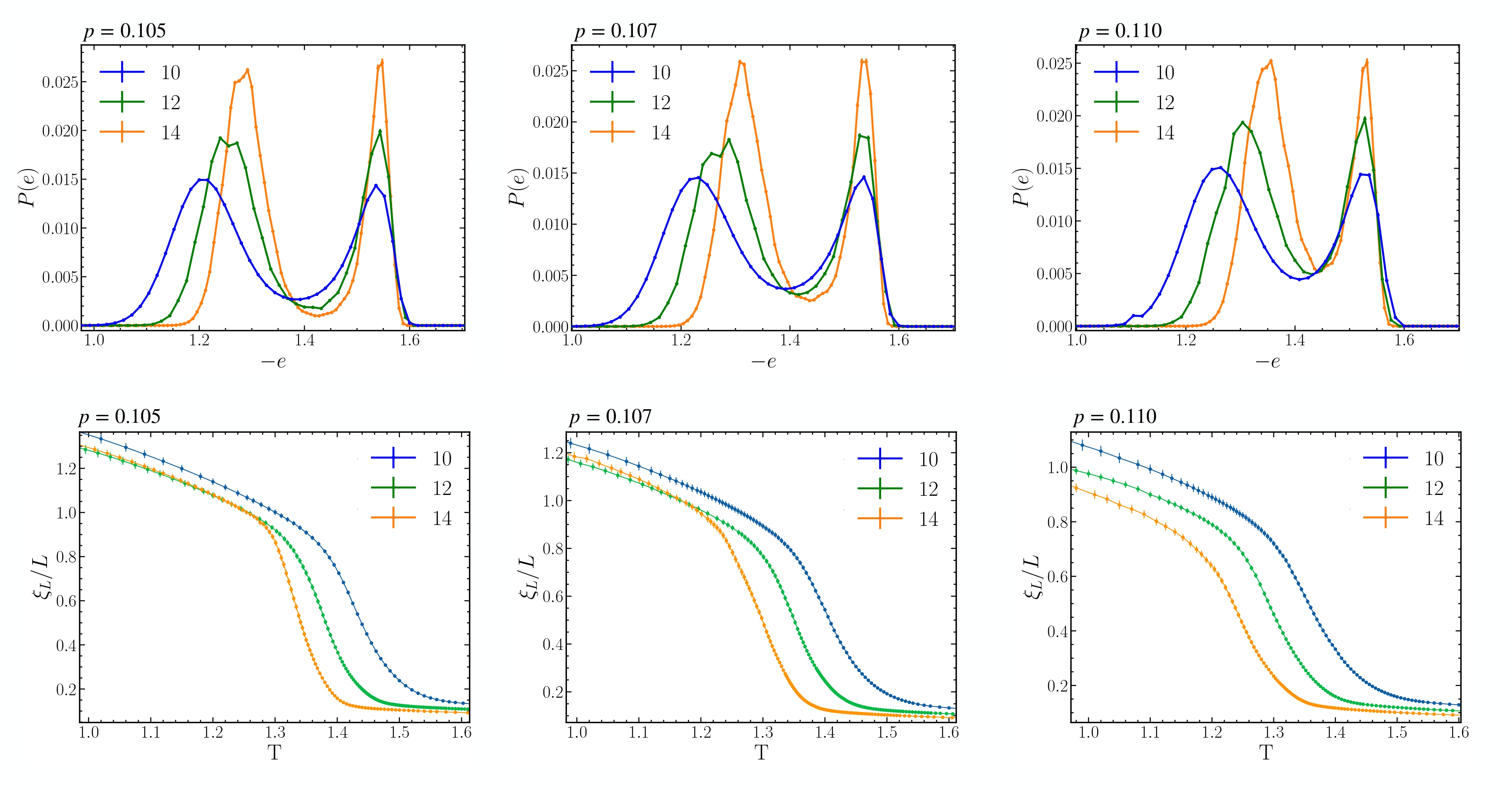}
  \caption{Disorder-averaged energy histograms (top row) and correlation lengths (bottom row) at different disorder values and for various system sizes. While the energy histograms present a typical first-order double-peak behavior at all disorder values due to strong finite-size effects, the presence of a first-order transition at $p=0.105$ is made obvious by the sharpening of the two peaks and a deepening of the minimum between them as the system size grows, a feature that can still be observed, although in less evident capacity, at $p=0.107$. This lowering of the minimum corresponds to a growing domain wall free energy cost as the system size increases which, in the thermodynamic limit, determines the presence of a first-order transition. On the other hand, at $p=0.110$ the minimum goes up with system size, indicating that the first-order features of the system are slowly disappearing. This is further highlighted by the sharp absence of any crossing in the correlation lengths regardless of system size, another indicator of a crossover regime. The sharp contrast between these behaviours allows us to identify the ordered and disordered phases of the model, and estimate the optimal error threshold to sit in between $0.107$ and $0.110$.}
  \label{fig:plots}
\end{figure*}

After taking into account the optimization of the temperature distribution, the system still required a remarkable amount of Monte Carlo sweeps to reach equilibration. The simulations were run on the LRZ KCS cluster \cite{LRZ_Linux_Cluster} and used over $7\times 10^6$ CPU hours in total, with an average of $\approx 10^6$ CPU hours of simulation time per $(L,p)$ dataset. Due to the cumbersome amount of computation time each point requires, computation was restricted to a limited set of $(L,p)$ points. The system sizes simulated lie in the range of $L\in[10,14]$, as lower system sizes were affected by excessively large finite-size effects which make them irrelevant for scaling purposes whereas higher system sizes were computationally unapproachable due to the exponential growth of the simulation times required. After a preliminary round of simulations used to infer a rough position of the error threshold, the range of relevant disorder values was originally set to be $p\in[0.105,0.110]$. To achieve an estimate of the threshold value with precision comparable to that of previous works \cite{Katzgraber09,Kubica18,Song22}, it was sufficient to simulate only 3 datapoints: the lower bound, an upper bound, and a conservative estimate of the possible location of the threshold, which was set at $p^{\text{mid}}=0.107$. Simulating additional intermediate disorder values would only increase computational cost without improving the statistical robustness of the threshold estimate: for this reason, the reported simulation set represents the minimal dataset that captures all qualitative changes relevant for locating the tricritical region while keeping the computational effort within feasible limits.

\begin{table}[tb]
\renewcommand{\arraystretch}{1.}
\centering
\begin{tabular}{c@{\hspace{1cm}}c@{\hspace{1cm}}c@{\hspace{1cm}}c@{\hspace{1cm}}c@{\hspace{1cm}}}
\toprule
$p$ & $L$ & $N_{d}$ & $N_{T}$ & $\tau$ \\
\midrule
$0.090$ & $12$ & $200$ & $96$ & $21$\\
\vspace{0.2cm}
$0.090$ & $14$ & $200$ & $96$ & $21$\\
$0.105$ & $10$ & $500$ & $64$ & $20$\\
$0.105$ & $12$ & $500$ & $128$& $22$\\
\vspace{0.2cm}
$0.105$ & $14$ & $500$ & $96$ & $22$\\
$0.107$ & $10$ & $500$ & $96$ & $21$\\
$0.107$ & $12$ & $500$ & $128$& $22$\\
\vspace{0.2cm}
$0.107$ & $14$ & $500$ & $128$& $23$\\
$0.110$ & $10$ & $500$ & $96$ & $22$\\
$0.110$ & $12$ & $500$ & $96$ & $23$\\
$0.110$ & $14$ & $500$ & $96$ & $23$\\
\bottomrule
\end{tabular}
\caption{Simulation parameters of the data shown in figures \ref{fig:090},\ref{fig:plots} for the Random-bond checkerboard code. $N_d$ denotes the number of independent simulations used for the disorder averaging of the respective system size $L$ at disorder value $p$. For each of these, $N_T$ different temperatures were simulated in parallel at different temperatures with a $PT$ update between adjacent replicas proposed, on average, every $10$ sweeps of Metropolis updates through the whole lattice. $\tau$ indicates the number of sweeps used for equilibration, with $2^{\tau}$ representing the disorder averaged equilibration time. To ensure proper equilibration and sampling for the majority of the replicas, each disorder simulation was run for approximately $2^{\tau+1}$ sweeps: $2^{\tau}$ for thermalization, followed by $2^{\tau}$ for sampling.}
\label{tab:simparams}
\end{table}

\subsection{Numerical Results}
We simulated the random Tetrahedral Ising model using large-scale parallel tempering Monte Carlo simulations, averaging the results of enough disorder configurations in order to obtain the disorder averaged features of the model. Each configuration simulated for a given value of the disorder parameter $p$ is obtained by starting from a fully FM set of couplings $(\eta_{\mathbf{v}}^{\pm}=+1 \,\,\forall\,\, \mathbf{v})$ and then proposing a random flip to each coupling with probability $p$. This way, the average ratio between number of AFM couplings and the total number of couplings equals $p$.

Accompanying the phenomenon of subsystem symmetry breaking~\cite{ Canossa23,Canossa24}, the model shows strong features of a first-order phase transition even in proximity of the error threshold. This property, coupled with the restricted set of lattice sizes accessible, requires us to carefully analyze the subtle changes in the thermodynamic features of the model between the system sizes available.
To infer whether the system shows a transition at a given disorder value, we studied the behavior of the energy histogram 
\begin{equation}
P(e) = \langle\delta\left(e - e^{\prime}\right)\rangle,
\end{equation}
where $e'=E'/N$ is the energy density for a given configuration of a system of size $N=L^3/2$,
and the second-moment correlation length 
\begin{equation}\label{eq:corrlength}
\xi_L = \frac{1}{2 \sin\left(|\textbf{k}_{\text{min}}/2|\right)} \left( \frac{\tilde{G}(\mathbf{0})}{\tilde{G}(\textbf{k}_{\text{min}})} - 1 \right)^{1/2}.
\end{equation}
Here, $\tilde{G}(\textbf{k}):= \sum_\textbf{r} G(\textbf{r})e^{-i\textbf{k}\cdot\textbf{r}}$ is the Fourier transform of the spatial correlator $G(\textbf{r})$ and $\textbf{k}_{\text{min}}$ is the smallest non-zero wave vector in reciprocal space~\cite{Janke08}. The spatial correlator of the model is defined as~\cite{Canossa24}
\begin{equation}\label{eq:correlator}
G(r) = \frac{1}{N} \sum_{\bf v} \mean{\sigma_{\bf v} \sigma_{{\bf v} + \hat{x} + \hat{y}} \sigma_{{\bf v} + \hat{y} + r\hat{z}} \sigma_{{\bf v} + \hat{x} + r\hat{z}}}.
\end{equation}




The first-order nature of the phase transitions is primarily identified by the characteristic sharpening of the double peaks present in the energy histograms: as the system size grows, the well between the two peaks should deepen, denoting an increase in the latent heat required to break the phase boundaries (see Fig.~\ref{fig:090}). On the other hand, the absence of a clear phase transition between ordered and disordered phase in the thermodynamic limit is diagnosed by observing that the two peaks tend to get closer and less defined as the system size grows larger, meaning that the presence of two peaks is only a finite-size effect and that, in the thermodynamic limit, these will eventually collapse into a single peak.
To corroborate this evidence, the second-order correlation length is used: while in the case of first-order phase transitions this is much more susceptible to finite-size effects and is not a good indicator of the transition point, this can still be used to discriminate between a disorder-induced crossover regime, where the correlation length should not show any crossing, a first-order phase transition, where a crossing is not expected but may still occur due to finite-size effects, and a second-order phase transition, where a crossing is expected in the large-system limit.

In order to prove whether the checkerboard code saturates the threshold bound, the search for the error threshold was done at first for disorder values around $p = 0.11$, the expected critical point for self-dual models with quenched disorder~\cite{Takeda05,Nishimori07,Song22}.
At $p=0.110$ and above, the double peak behavior in the energy histograms is still visible, but tends to disappear at larger system sizes. This, combined with the lack of crossing in the correlation length (which rules out the possible presence of a continuous phase transition), confirms that the system is indeed in a crossover regime and that the double-peak behavior observed is caused by finite-size effects and will disappear in the thermodynamic limit. This serves as a characteristic example of the behavior of the system in the disordered regime and a reference against which to compare observations at lower error rates, especially due to the fact that the decisive signatures lie in the ways in which the strong finite-size effects evolve as the system size increases.

At disorder value $p=0.105$ and below, the two peaks tend to sharpen and the gap between them deepens as the system size increases, indicating the presence of a first-order phase transition. Given the first-order nature shown by the energy histograms, a crossing in the correlation length is not necessarily expected. Nevertheless, the observation of a crossing between the larger system sizes is consistent with the relaxation of finite-size effects and rules out the eventuality of a crossover regime.

The behavior shown at $p=0.107$ was similar, although much less accentuated, as the minimal point in the gap of the energy histograms lowers only slightly with system sizes; considering the presence of a crossing in the correlation lengths between the two curves at larger system sizes, we argue that the system is still in the decodable phase, but is in very close proximity to the threshold. 

These observations allow us to estimate the tricritical point. Since we can place it within the interval $[0.105,0.110)$, the most conservative claim that does not exclude any value in this range is to set it at $p_{th} \simeq 0.107(3)$. The size of the error bar is comparable to that in previous numerical threshold estimates (see Table~\ref{tab:threshold}) and is sufficient to cover the full interval of uncertainty while remaining deliberately conservative.

This threshold represents the highest optimal error threshold among any $3D$ topological codes.
Moreover, it gives to $H(p_{th}^X) + H(p_{th}^Z) \approx 0.98(2)$, which nearly saturates the error threshold bound and also confirms the applicability of the generalized duality for the checkerboard code.

\begin{figure}[!t]
\centering
\includegraphics[width =.48\textwidth]{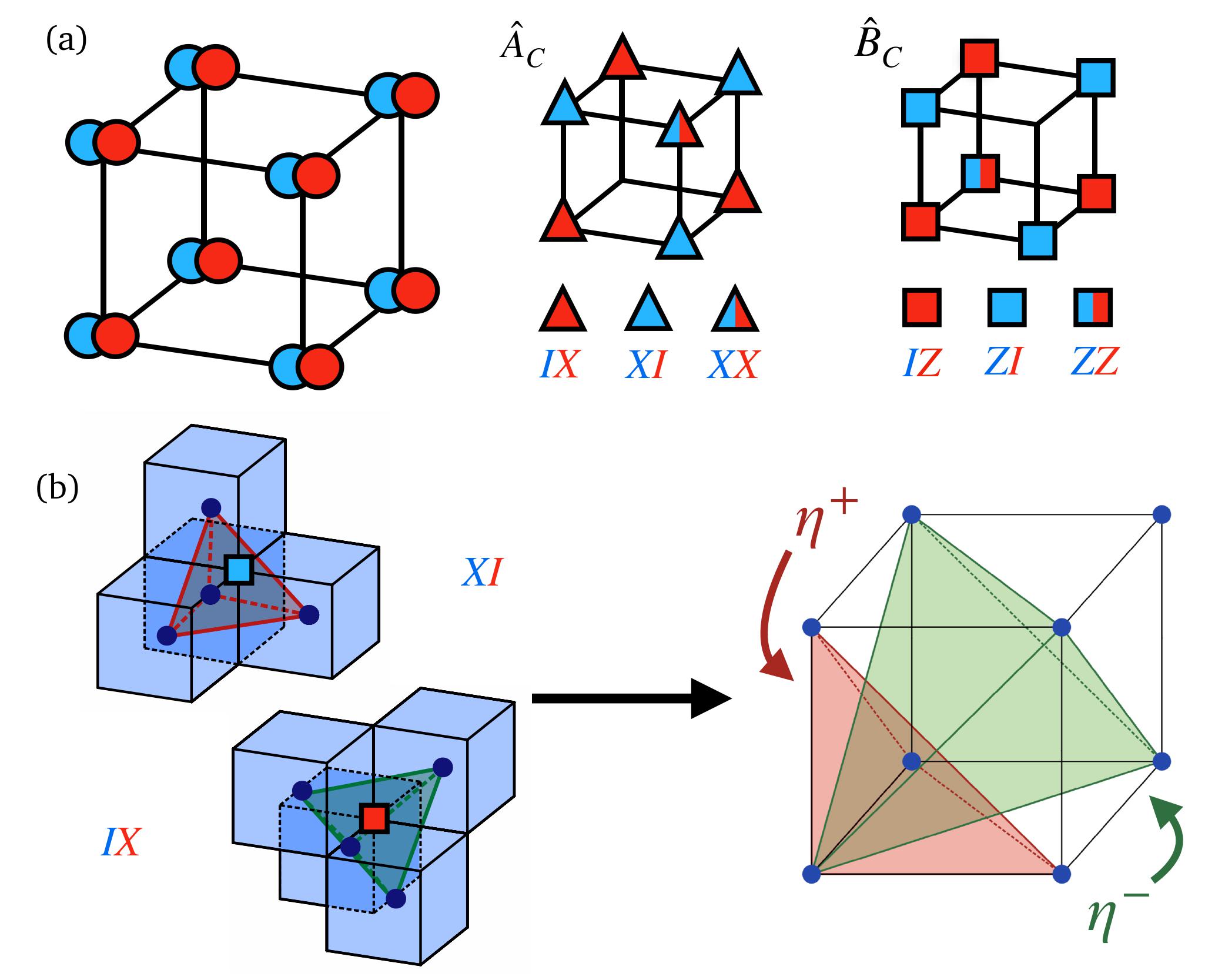}
\caption{(a) Illustration of the stabilizer structure of the Haah's code. Two sets qubits are situated at the vertices of a cubic lattice. The $X-$ and $Z-$ type stabilizers $A_C,B_C$ are defined for each unit cell $C$ as a product of Pauli operators acting on a subset of the 16 qubits residing at the vertices of the cube. (b) A single-qubit $X$ (or $Z$) error excites 4 cubic stabilizer operators: since the structures are equivalent for $X$ and $Z$ stabilizers up to a global rotation, both independent noise models can be represented by the random Fractal Ising Model Eq.~\eqref{eq:FIM_model}.}
\label{fig:Haah's_model}
\end{figure}

\section{Considerations about Haah's code}\label{sec:Haah}
Haah’s cubic code is a seminal fracton code, notable for having no string-like logical operators and for being one of the first constructions to reveal fracton order in a quantum error-correction context~\cite{Haah11}. The code is defined on a cubic lattice with two qubits per site, and its stabilizers follow the general structure introduced in Eq.~\eqref{eq:Check_Ham}: each cube hosts an $X-$ and $Z-$type 8-body stabilizer. Identifying each cube $C$ with its $(0,0,0)$ vertex $\bf v$, and labelling the two families of qubits as $\{1,2\}$, the stabilizers take the form
\begin{flalign}
\hat{A}_{C} &=  \hat{\sigma}^{X,2}_{\mathbf{v} + \hat{x}} 
                \,\hat{\sigma}^{X,2}_{\mathbf{v} + \hat{x}}
                \,\hat{\sigma}^{X,1}_{\mathbf{v} + \hat{x}+ \hat{y}}
                \,\hat{\sigma}^{X,2}_{\mathbf{v} + \hat{z}}\,\,\,
                \,\hat{\sigma}^{X,1}_{\mathbf{v} + \hat{z}+ \hat{x}}
                \,\hat{\sigma}^{X,1}_{\mathbf{v} + \hat{z}+ \hat{y}}
                \,\hat{\sigma}^{X,1}_{\mathbf{v} + \hat{x}+ \hat{y}+ \hat{z}}
                \,\hat{\sigma}^{X,2}_{\mathbf{v} + \hat{x}+ \hat{y}+ \hat{z}} \nonumber ,\\
\hat{B}_{C} &=  \hat{\sigma}^{Z,1}_{\mathbf{v}}\,\,
                \,\hat{\sigma}^{Z,2}_{\mathbf{v}}\,\, 
                \,\hat{\sigma}^{Z,2}_{\mathbf{v} +\hat{x}}\,\,\,
                \,\hat{\sigma}^{Z,2}_{\mathbf{v} +\hat{x}+\hat{y}}
                \,\hat{\sigma}^{Z,1}_{\mathbf{v} +\hat{x}+\hat{y}}
                \,\hat{\sigma}^{Z,2}_{\mathbf{v} +\hat{z}}\,\,\,
                \,\hat{\sigma}^{Z,1}_{\mathbf{v} +\hat{z}+\hat{x}}\,\,\,\,\,
                \,\hat{\sigma}^{Z,1}_{\mathbf{v} +\hat{z}+\hat{y}} \nonumber.
\end{flalign}

The model is characterized by a subextensive number of fractal symmetries ~\cite{Haah13}: as a consequence of these $\mathbb{Z}_2$ fractal conservation law, a single-qubit error creates a cluster of four fracton excitations which cannot be separated by any local operator. These defects can only move collectively along the underlying fractal pattern and are therefore individually immobile.

The noise model of the Haah's code for independent $X$ and $Z$ errors can be represented through the SM mapping with a classical Ising model with 4-body interactions. Taking into account the fact that each vertex contains two qubits, 
The Hamiltonian consists of two four-body interactions on the tetrahedra as depicted in Fig.~\ref{fig:Haah's_model}: this leads to a random Fractal Ising model
\begin{flalign} \label{eq:FIM_model}
\mathcal{H}_{\rm FIM} = -\sum_{\mathbf{v}}\Big( \,\,& \eta^{+}_{\mathbf{v}}\,\sigma_{\mathbf{v}}\sigma_{\mathbf{v}+\hat{x}}\sigma_{\mathbf{v}+\hat{y}}\sigma_{\mathbf{v}+\hat{z}} \nonumber \\ +  & \eta^{-}_{\mathbf{v}}\,\sigma_{\mathbf{v}}\sigma_{\mathbf{v}+\hat{x}+\hat{y}}\sigma_{\mathbf{v}+\hat{y}+\hat{z}}\sigma_{\mathbf{v}+\hat{x}+\hat{z}} \,\,\Big),
\end{flalign}
For simplicity, we take $\eta^+ = \eta^- = 1$, meaning that the two qubit families have the same error probability $p$.

The error-threshold analysis of the model was found to be numerically intractable due to the non-trivial nature of the fractal subsystem symmetries. In presence of periodic boundary conditions, the way in which the fractal excitation propagates gives ground state degeneracies which strictly depend on the system size $L$:\begin{align}\label{eq:GSD}
\text{GSD}(H_{\mathrel{\text{FIM}}}) = 
\begin{cases}
2^{2L-1}, & L=2^{n},\\
2^{2L-5}, & L=4^{n}-1,\\
2^{2^m-1}, & L= 2^{m-1}(2^{n}+1), \\
2^{2^m-1}, & L= 2^{m-1}(2^{2n-1}-1), \\
\end{cases}
\end{align}
with integers $n, m \geq1$~\cite{Canossa24}. 
This strongly affects the behavior of the system, leading to different finite-size effects for the different system size sequences.
Even for the simplest size sequence $L=2^n$, the accessible lattice sizes are restricted to $L = 4, 8$ which are insufficient to locate the phase transition due to strong finite-size effects.
Larger systems ($L \geq 16$) are computationally prohibitive, with preliminary runtime estimates approaching $\approx 1.5\times10^7$ additional CPU hours of computation per $(L=16,p)$ dataset.

Nevertheless, given the extensive justification of the generalized entropy duality Eq.~\eqref{eq:duality} for standard topological codes and fracton codes, it is reasonable to conjecture Haah’s code to behave similarly. 
Since the Haah's code, we can expect that its code capacity threshold should lie very close to the $p_{th}\approx 0.11$ bound, even though a direct numerical confirmation remains an open challenge.

\section{Summary} \label{sec:sum}

In this work, we investigated the fault-tolerance threshold of self-dual fracton codes. By mapping the error correction problem onto a statistical mechanical model and performing large-scale numerical simulations, we computed the optimal thresholds of the checkerboard code against stochastic Pauli noise. Our numerical analysis identifies a code capacity threshold of $p_{th} \simeq 0.107(3)$ for the checkerboard code. This value is notable not only for its magnitude—outperforming other $3D$ codes—but because it represents the first instance of a $3D$ code that nearly saturates the theoretical limit of $p_{th} \approx 0.11$ allowed for topological stabilizer codes~\cite{Dennis02}.

Moreover, our results further corroborate the generalized entropy duality in Eq.\eqref{eq:duality}, confirming its applicability beyond standard topological codes.
The remarkable effectiveness of this duality may be attributed to the self-averaging nature of both the ordered and disordered paramagnetic phases and the lack of spin glass phase along the Nishimori line. This ensures that the self-consistent replica analysis yields reasonable descriptions.
Our validation carries significant practical importance alongside its theoretical implications.
Exact threshold computations are notoriously resource-intensive and generally lack efficient algorithms. For $3D$ codes in particular, determining their thresholds can consume millions of CPU hours. The generalized duality can save immense computational resources. Leveraging this predictive power, we conjecture that the code capacity threshold of Haah’s code, which is even more computationally demanding to simulate, also approaches the theoretical limit of $p_{th} \approx 0.11$.
Similarly, one can also expect the bivariate bicycle codes~\cite{Bravyi24}, which are a rich class quantum low-density parity-check codes characterized by symmetry-enriched topological order~\cite{Chen25}, also possess such an optimal error threshold.

Looking forward, while this work focused on code capacity threshold, a critical next step is to extend the analysis to include measurement errors and circuit-level noise. Developing effective strategies to treat these more general noise models beyond the surface code remains an exciting open question. Furthermore, from a theoretical perspective, it is of great interest to investigate the precise nature of the correctable-to-uncorrectable phase transition across the threshold, which may reveal novel critical behaviors unique to fracton states of matter.

\emph{Acknowledgments}. G.C., K.L., and L.P. acknowledge support from the Deutsche Forschungsgemeinschaft (DFG, German Research Foundation) under Germany's Excellence Strategy -- EXC-2111 -- 390814868.
MAMD acknowledges support from Spanish MICIN grant PID2021-122547NB-I00 and the ``MADQuantumCM'' project funded by Comunidad de Madrid (Programa de acciones complementarias) and by the Ministry for Digital Transformation and of Civil Service of the Spanish Government through the QUANTUM ENIA project call –Quantum Spain project, and by the European Union through the Recovery, Transformation and Resilience Plan Next Generation EU within the framework of the Digital Spain 2026 Agenda, the CAM Programa TEC-2024/COM-84 QUITEMAD-CM.
H.S. acknowledges support from the National Natural Science Foundation of China (Grants No.~12522502, No.~12474145, 
and No.~12447101).
L.P. acknowledges financial support from ANR-23-CE30-0018.
The project/research is part of the Munich Quantum Valley, which is supported by the Bavarian state government with funds from the Hightech Agenda Bayern Plus.

\emph{Data availability}. The Monte Carlo simulation data supporting the findings of this study are available in the repository indicated in Ref.~\cite{canossa_2025_repo}.

\bibliography{checkerboard.bib}

\end{document}